\documentclass[a4paper,10pt]{article}
\usepackage{theorem}
\usepackage{latexsym,amssymb,amsfonts,amsmath,cite}
\usepackage[dvips]{graphicx}
\newtheorem{thm}{Theorem}
\newtheorem{lem}{Lemma}
\newtheorem{cor}[thm]{Corollary}
\newtheorem{pro}[thm]{Proposition}

\newcommand{\qed}{\hfill $\Box$}

\begin{document}
\title{{\Large {\bf Uniform stationary measure of space-inhomogeneous \\ quantum walks in one dimension}}}
\author{
{\small Yusuke Ide}\\
{\scriptsize School of Knowledge Science, 
Japan Advanced Institute of Science and Technology}\\
{\scriptsize Nomi 923-1292, Japan}\\
{\scriptsize e-mail: ide@jaist.ac.jp}\\
\\
{\small Norio Konno}\\
{\scriptsize Department of Applied Mathematics, 
Faculty of Engineering, 
Yokohama National University}\\
{\scriptsize Hodogaya, Yokohama 240-8501, Japan}\\
{\scriptsize e-mail: konno-norio-bt@ynu.ac.jp}\\
\\
{\small Daichi Nakayama}
\footnote{To whom correspondence should be addressed. E-mail: nakayama-daichi-gw@ynu.jp}\\
{\scriptsize Department of Applied Mathematics, 
Faculty of Engineering, 
Yokohama National University}\\
{\scriptsize Hodogaya, Yokohama 240-8501, Japan}\\
{\scriptsize e-mail: nakayama-daichi-gw@ynu.jp}\\
}
\vskip 1cm

\date{\today }
\pagestyle{plain}

\maketitle

\par\noindent
\begin{small}

\par\noindent
\begin{abstract}
The discrete-time quantum walk (QW) is a quantum version of the random walk (RW) and 
has been widely investigated for the last two decades. Some remarkable properties of QW are well known. 
For example, 
QW has a ballistic spreading, i.e., QW is quadratically faster than RW.
For some cases, localization occurs: a walker stays at the starting position forever.
In this paper, we consider stationary measures of two-state QWs on the line.
It was shown that for any space-homogeneous model, 
the uniform measure becomes the stationary measure.
However, the corresponding result for space-inhomogeneous model is not known.
Here, we present a class of space-inhomogeneous QWs on the line and cycles in which the uniform measure is stationary.
Furthermore, we briefly discuss a difference between QWs and RWs.
\end{abstract}

\end{small}

\section{Introduction} 
The discrete-time quantum walk (QW) is a quantum analog of the random walk 
(RW) \cite{AharonovEtAl1993, AmbainisEtAl2001, Gudder1988, MeyerEtAl1996}. 
There are two well-known important properties for QWs. 
The first one is ballistic spreading, i.e., the standard deviation of 
the walker's position grows linearly in time, 
quadratically faster than RW \cite{Konno2002, Konno2005}. 
The second one is localization, i.e., a walker stays at the starting position forever. 
In some cases, QWs have both ballistic spreading and localization \cite{Konno2010}.

There are not only above-mentioned theoretical aspects but also practical application of QW, such as the strongly correlated electron system
\cite {OkaEtAl2005}, topological insulators \cite {Kitagawa2012, ObuseEtAl2011, ObuseEtAl2015}, and 
radioactive waste reduction \cite{IchiharaEtAl2013, MatsuokaEtAl2011, MatsuokaEtAl2013}. 
For more detailed information on QWs, see 
Venegas-Andraca \cite{Venegas2008, Venegas2012}, Konno \cite{Konno2008b}, Cantero et al. \cite{CanteroEtAl2013}, Manouchehri and Wang \cite{MW2014}, Portugal \cite{P2013}.

Stationary measures of discrete-time QWs have been extensively investigated from around 2014
\cite {
ek2014,
EndoEtAl2016,
EndoEtAl20162,
Kawai2017,
Kawai2017_2,
Komatsu2017,
Konno2014,
kt2015,
WangEtAl2015}.
In this paper, we focus on stationary measures of two-state QWs in one dimension. Konno \cite{Konno2014} proved that the uniform measure becomes the stationary measure for any space-homogeneous model. We call this measure ``uniform stationary measure''. However, the corresponding result for space-inhomogeneous model is not known. This is a motivation of our study. Here, we present a class of space-inhomogeneous QWs on the line and cycles in which the uniform measure is stationary. See Theorem 3.2 for line case and Proposition 4.1 for cycle case.

Moreover, we consider a difference between the periodicity of coins for the QW and the RW. As for our results, see Corollary 5.1 (QW case) and Proposition 6.1 (RW case).

The rest of this paper is organized as follows. 
In Section 2, we introduce the two-state space-inhomogeneous QW on the line. 
Section 3 presents our main result and the proof. Section 4 deals with an application to cycles. 
We consider the periodicity of coins for QW (Section 5) and RW (Section 6). 
Section 7 is devoted to summary.

\section{Model and method} 
We introduce a discrete-time space-inhomogeneous QW on the line which is a quantum version of the RW 
with an additional coin state. 
Let $\mathbb{N}$ be the set of all natural numbers, 
$\mathbb{Z}$ be the set of all integers and $\mathbb{C}$ be the set of all complex numbers.
The walker has a coin state, $\Psi_{n}(x)$, at time $n (\in \mathbb{N}$) and position $x (\in \mathbb{Z}$) described
by a two-dimensional vector:
\begin{align*}
\Psi_{n}(x) = 
\begin{bmatrix}
\Psi^{L}_{n}(x)\\
\Psi^{R}_{n}(x)\\
\end{bmatrix}
\in \mathbb{C}^{2}. 
\end{align*}
The upper and lower elements are referred as left and right chiralities, respectively. The time evolution is determined
by $2\times2$ unitary matrices $U_{x}$ which is called coin matrix here:
\begin{align*}
U_{x} = 
\begin{bmatrix}
a_{x} & b_{x}\\
c_{x} & d_{x}\\
\end{bmatrix}
\hspace{2.em}(x \in \mathbb{Z}) .
\end{align*}
The subscript $x$ stands for the position. We divide $U_{x}$ into $U_{x} = P_{x} + Q_{x}$ with
\begin{align*}
P_{x} = 
\begin{bmatrix}
a_{x} & b_{x}\\
0 & 0\\
\end{bmatrix}
,
Q_{x} =
\begin{bmatrix}
0 & 0\\
c_{x} & d_{x}\\
\end{bmatrix}.
\end{align*}
The $2 \times 2$ matrix $P_{x}$(resp. $Q_{x}$) represents the walker's movement to the left (resp. right) at position $x$ at
each time step. Then the time evolution of the walk is defined by
\begin{align*}
\Psi_{n+1}(x) \equiv P_{x+1} \Psi_{n}(x+1) + Q_{x-1} \Psi_{n}(x-1) .
\end{align*}
That is 
\begin{align*}
\begin{bmatrix}
\Psi^{L}_{n+1}(x)\\
\Psi^{R}_{n+1}(x)\\
\end{bmatrix}
=
\begin{bmatrix}
a_{x+1}\Psi^{L}_{n}(x+1) + b_{x+1}\Psi^{R}_{n}(x+1)\\
c_{x-1}\Psi^{L}_{n}(x-1) + d_{x-1}\Psi^{R}_{n}(x-1)\\
\end{bmatrix}
.
\end{align*}
Now let
\begin{align*}
\Psi_{n}
= {}^{T}[\hspace{.3em}...\hspace{.4em}, \Psi^{L}_{n}(-1),  \Psi^{R}_{n}(-1), \Psi^{L}_{n}(0),  \Psi^{R}_{n}(0),
 \Psi^{L}_{n}(+1),  \Psi^{R}_{n}(+1), \hspace{.4em}...\hspace{.3em}] , 
\end{align*}
\begin{align*}
U^{(s)}
= 
\left[ 
\begin{array}{ccccccc}
\ddots & \vdots & \vdots & \vdots & \vdots & \vdots & \vdots \\
\cdots & Q_{-2} & O & P_{0} & O & \cdots & \cdots \\
\cdots & O & Q_{-1}& O & P_{1} & O & \cdots\\
\cdots & \cdots & O & Q_{0} & O & P_{2} & \cdots\\
\cdots & \vdots & \vdots & \vdots & \vdots & \vdots & \ddots 
\end{array} 
\right] \hspace{.4em} with \hspace{.4em} O = \left[
\begin{array}{cc}
0 & 0 \\ 
0 & 0 \\ 
\end{array}
\right] ,
\end{align*}
where $T$ means the transposed operation. 
Then the state $\Psi_{n}$ of the QW at time $n$ is given by 
$\Psi_{n} = (U^{(s)})^{n} \Psi_{0}$ for any $n \geq 0$. 
Put $\mathbb{R}_{+} = [0, \infty)$. 
Here we introduce a map 
$\Gamma : (\mathbb{C}^2)^\mathbb{Z} \to \mathbb{R}_{+}^\mathbb{Z}$, 
as
\begin{align*}
\Gamma(\Psi) \equiv {}^T \bigg[\cdots, 
|\Psi^L (-1)|^{2} + |\Psi^R (-1)|^{2}, 
|\Psi^L (0)|^{2} + |\Psi^R (0)|^{2}, 
|\Psi^L (1)|^{2} + |\Psi^R (1)|^{2}, 
\cdots\bigg] \in \mathbb{R}_{+}^{\mathbb{Z} },
\end{align*}
for
\begin{align*}
\Psi={}^T\bigg[\cdots,
\begin{bmatrix}
\Psi^L (-1)\\
\Psi^R (-1)
\end{bmatrix},
\begin{bmatrix}
\Psi^L (0)\\
\Psi^R (0)
\end{bmatrix},
\begin{bmatrix}
\Psi^L (1)\\
\Psi^R (1)
\end{bmatrix},
\cdots\bigg]\in(\mathbb{C}^2)^{\mathbb{Z}}.
\end{align*}
\noindent
By using the map $\Gamma$, we define a measure $\mu:\mathbb{Z} \to \mathbb{R}_{+}$ by
\begin{align*}  
\mu(x) \equiv \Gamma(\Psi)(x)=|\Psi^L(x)|^2+|\Psi^R(x)|^2 \ \ \ (x\in \mathbb{Z}).
\end{align*}
Let $\mathcal{M}(U^{(s)})$ be the set of the measures of the QW.
Then, we put the set
\begin{equation*}
\mathcal{M}_s(U^{(s)})=\Big\{\mu\in\mathcal{M}(U^{(s)}): {}^{\exists} \Psi_0\in\left(\mathbb{C}^2\right)^{\mathbb{Z}} \hspace{.2em}
s.t. \hspace{.4em}
\Gamma{((U^{(s)}})^n\Psi_0)=\mu\ (n=0,1,2,\ldots)\Big\}.
\end{equation*}
\noindent
We call the element of $\mathcal{M}_s(U^{(s)})$ the stationary measure of the QW.
Generally, the set of stationary measures of QW depends on its time evolution operator. 
It was shown by \cite{Konno2014} that two-state space-homogeneous QWs in one dimension 
have a uniform stationary measure.
In space-homogeneous case, the coin matrix $U_{x}$ does not depend on $x$.
In order to emphasize the homogeneity, we use $U^{(s)}_{hom}$ as $U^{(s)}$. 
Thus, if we define $\mathcal{M}_{unif}$
be the set of uniform measures on $\mathbb{Z}$; 
\begin{equation*}
\mathcal{M}_{unif} = \Big\{\mbox{measure} \hspace{.5em}\mu : \hspace{.2em} 
\mbox{there exists}  \hspace{.5 em} c > 0 \hspace{.5 em} \mbox{such that} 
\hspace{.5 em} \mu (x) = c \hspace{.5 em} \mbox{for any} \hspace{.5 em} x \in \mathbb{Z}\Big\},
\end{equation*}
then we can write $\mathcal{M}_{unif} \subset \mathcal{M}_s(U^{(s)}_{hom})$.

On the other hand, it was not known whether 
a two-state space-inhomogeneous QW in one dimension
has a uniform measure as a stationary measure or not. 
Let us consider the eigenvalue problem
\begin{align*}  
U^{(s)} \Psi = \lambda \Psi,
\end{align*}  
where $\lambda \in \mathbb{C}$ with $|\lambda| = 1$.
If we assume that the initial state is $\Psi_{0}$ satisfying $U^{(s)} \Psi_{0} = \lambda \Psi_{0}$,
then we have
\begin{align*}
\Psi_{n} = (U^{(s)})^{n} \Psi_{0} = \lambda^{n} \Psi_{0}.
\end{align*}
Noting that 
$|\lambda| = 1$, we see that for any $n \geq 0$,
\begin{align*} 
\mu_{n}(x) = \Gamma(\Psi_{n})(x)  
= \Gamma(\lambda^{n} \Psi_{0})(x)
= |\lambda^{n} |^{2}  \Gamma(\Psi_{0})(x) 
=  \Gamma(\Psi_{0})(x) = \mu_{0}(x).
\end{align*}
This shows that $\mu_{0}$ is a stationary measure, i.e., $\mu_{0} \in \mathcal{M}_s(U^{(s)})$. 
For $\lambda \in \mathbb{C}$ with $|\lambda|=1$, we put 
\begin{align*}
\mathcal{A}_{e}^{(\lambda)}(U^{(s)})=\Big\{\Psi\in \left(\mathbb{C}^{2}\right)^{\mathbb{Z}} \setminus \{0\} :\hspace{.2em}
U^{(s)} \Psi = \lambda \Psi
 \Big\}.
\end{align*}
 Moreover, we introduce the following set of measures: 
tIf we define the following set of measures:
\begin{align*}
\mathcal{M}_{s}^{e}(U^{(s)})=
\bigcup_{\lambda \in \mathbb{C}, |\lambda| = 1} \mathcal{M}_{e}^{(\lambda)}(U^{(s)})
\end{align*}
Clearly, we obtain $\mathcal{M}_{s}^{e} (U^{(s)}) \subset \mathcal{M}_{s} (U^{(s)})$.

\section{Main result} 
In this section, we consider QWs defined by the parameters $\{ \omega_{x} \}_{x \in \mathbb{Z}}$. 
The coin at position $x$ of the QWs is 
\begin{align*}
U_{x} = 
\left[ 
\begin{array}{cc}
\cos\theta & e^{i \omega_{x}}\sin \theta  \\
e^{-i\omega_{x}}\sin \theta & -\cos \theta  \\ 
\end{array} 
\right] \hspace{2. em} (\omega_{x} \in [0, 2\pi), \theta \in (0, 2\pi)).
\end{align*}
In particular, we focus on a class of QWs satisfying that there exists $\phi \in [0, 2\pi)$ such that 
$\omega_{x} - \omega_{x-1} = 2\phi \hspace{.2em} (\hspace{-.6em} \mod 2\pi )$ for all $x \in \mathbb{Z}$.
We write this class of QWs defined by the sequence of coin $ \{U_{x}\}_{x \in \mathbb{Z}} $ as $\mathcal{C}_{\phi}$ and 
simply denote $U_{x} \in \mathcal{C}_{\phi}$. 
Before showing our result, we introduce the following result in \cite{Kawai2017_2}.
\begin{lem} ( Theorem 3.1 in \cite{Kawai2017_2} ) \\  
Let $\Psi(x) = {}^{T}[\Psi^{L}(x), \Psi^{R}(x)]$ be the amplitude. 
The coin matrix is defined by
\begin{align*}
U_{x} = 
\left[ 
\begin{array}{cc}
a_{x} & b_{x} \\
c_{x} & d_{x} \\ 
\end{array} 
\right]
\hspace{2. em} (x \in \mathbb{Z})
\end{align*}
where
$a_{x}b_{x}c_{x}d_{x} \neq 0$.
Then a solution of the following eigenvalue problem:
\begin{align*}
U^{(s)} \Psi = \lambda \Psi
\end{align*}
is given by
\begin{align}
\Psi(x)=
\begin{cases}
\displaystyle \prod_{y=1}^{x} D^{+}_{y} \Psi(0)&(x\geq 1),\\
\Psi(0)&(x=0),\\
\displaystyle \prod_{y=-1}^{x} D^{-}_{y} \Psi(0)&(x\leq -1).
\end{cases}
\end{align}
where
\begin{align*}
D^{+}_{x}=\begin{bmatrix}
\dfrac{\lambda^{2}-b_x c_{x-1}}{\lambda a_x}&-\dfrac{b_x d_{x-1}}{\lambda a_x}\\
\dfrac{c_{x-1}}{\lambda}&\dfrac{d_{x-1}}{\lambda}
\end{bmatrix},\ \ 
D^{-}_{x}=\begin{bmatrix}
\dfrac{a_{x+1}}{\lambda}&\dfrac{b_{x+1}}{\lambda}\\
-\dfrac{a_{x+1}c_x}{\lambda d_x}&\dfrac{\lambda^2-b_{x+1} c_{x}}{\lambda d_x}
\end{bmatrix}.
\end{align*}
These matrices both $D^{+}_{x}$ and $D^{-}_{x}$ are called transfer matrices. 
Moreover, we define $\{ \Gamma(\Psi) \}$ as the set of all $\Gamma(\Psi)$ given by Theorem 3.1 in \cite{Kawai2017_2}.
Therefore, 
\begin{align*}
\{ \Gamma(\Psi) \} = \mathcal{M}_{s}^{e} (U^{(s)}) (\subset \mathcal{M}_{s} (U^{(s)}) ).
\end{align*}
\end{lem}
\noindent
If we apply this result into our model, then we can get the following theorem. 
\begin{thm}
We consider QWs with $U_{x} \in \mathcal{C}_{\phi}$($\phi \in [0, 2\pi)$) and $\theta \neq \frac{\pi}{2}$, $\theta \neq \frac{3\pi}{2}$. 
Then we can construct $\Psi \in (\mathbb{C}^2)^\mathbb{Z}$ satisfying $U^{(s)} \Psi = e^{i \phi} \Psi$
in the following: 
\begin{align*}
\Psi(x)=
\begin{cases}
\displaystyle \prod_{y=1}^{x} D^{+}_{y} \Psi(0)&(x\geq 1),\\
\Psi(0)&(x=0),\\
\displaystyle \prod_{y=-1}^{x} D^{-}_{y} \Psi(0)&(x\leq -1),
\end{cases}
\end{align*}
where
\begin{align*}
D^{+}_{x}&=\begin{bmatrix}
e^{i\phi}\cos \theta & e^{i\alpha_{x}}\sin \theta \\
e^{-i\alpha_{x}}\sin \theta & -e^{-i\phi}\cos \theta 
\end{bmatrix} ,\ \ 
D^{-}_{x}=\begin{bmatrix}
e^{-i\phi}\cos \theta & e^{i\alpha_{x+1}}\sin \theta \\
e^{-i\alpha_{x+1}}\sin \theta & -e^{i\phi}\cos \theta 
\end{bmatrix}, \\
\alpha_{x} &= \phi + \omega_{x-1} = \omega_{x}-\phi \hspace{2.em}(\hspace{-.9em} \mod\hspace{.2em}2\pi).
\end{align*}
Moreover, $\Gamma (\Psi) \in \mathcal{M}_{unif} \cap \mathcal{M}_{s}^{e} (U^{(s)})\hspace{0.5em}
(\hspace{0.3em} \subset \mathcal{M}_{s} (U^{(s)})\hspace{0.3em})$. That is, our QW model has a uniform stationary measure.
\end{thm}
{\bf Proof.} 
In the case of QWs with $U_{x} \in \mathcal{C}_{\phi}$ ($\phi \in [0, 2\pi)$), 
considering the following eigenvalue problem
\begin{align*}
U^{(s)} \Psi = e^{i\phi} \Psi,
\end{align*}
then we calculate the transfer matrices as follows.
\begin{align*}
D_{x}^{+} 
&=
\begin{bmatrix}
\dfrac{\lambda^{2}-b_{x} c_{x-1}}{\lambda a_x}&-\dfrac{b_x d_{x-1}}{\lambda a_x}\\
\dfrac{c_{x-1}}{\lambda}&\dfrac{d_{x-1}}{\lambda}
\end{bmatrix} \\
&=
\begin{bmatrix}
\dfrac{e^{2i\phi} - e^{i\omega_{x}}\sin \theta e^{-i\omega_{x-1}}\sin \theta}{e^{i\phi}\cos \theta}&
-\dfrac{e^{i\omega_{x}}\sin \theta (- \cos \theta)}{e^{i\phi} \cos \theta}\\
\dfrac{e^{-i\omega_{x-1}}\sin \theta}{e^{i\phi}}&
-\dfrac{\cos \theta}{e^{i\phi}}
\end{bmatrix} \\
&= 
\begin{bmatrix}
\dfrac{e^{i\phi} - e^{i (\omega_{x} - \omega_{x-1} - \phi)} \sin^{2} \theta}{\cos \theta}&
e^{i (\omega_{x} - \phi) }\sin \theta\\
e^{-i (\omega_{x-1} + \phi) } \sin \theta&
- e^{-i\phi} \cos \theta
\end{bmatrix} \\
&= 
\begin{bmatrix}
e^{i\phi}\cos \theta & e^{i\alpha_{x}}\sin \theta \\
e^{-i\alpha_{x}}\sin \theta & -e^{-i\phi}\cos \theta 
\end{bmatrix}. 
\end{align*}
\noindent
The forth equality comes from the assumption that 
$\omega_{x} - \omega_{x-1} = 2\phi \hspace{.2em} (\hspace{-.6em} \mod 2\pi )$ for all $x \in \mathbb{Z}$.
Because of the assumption, we get 
$\phi + \omega_{x-1} = \omega_{x}-\phi \hspace{.2em}(\hspace{-.6em} \mod\hspace{.2em}2\pi)$ for all $x \in \mathbb{Z}$.
We write
$\alpha_{x} = \phi + \omega_{x-1} = \omega_{x}-\phi \hspace{.2em}(\hspace{-.6em} \mod\hspace{.2em}2\pi)$. 
As in the case of $D_{x}^{+}$, we compute $D_{x}^{-}$ in the following. 
\begin{align*}
D_{x}^{-} 
&=
\begin{bmatrix}
\dfrac{a_{x+1}}{\lambda}&\dfrac{b_{x+1}}{\lambda}\\
-\dfrac{a_{x+1}c_x}{\lambda d_x}&\dfrac{\lambda^2-b_{x+1} c_{x}}{\lambda d_x}
\end{bmatrix} \\
&=
\begin{bmatrix}
\dfrac{\cos \theta}{e^{i\phi}}&
\dfrac{e^{i\omega_{x+1}}\sin \theta}{e^{i\phi}}\\
-\dfrac{e^{-i\omega_{x}}\sin \theta \cos \theta}{-e^{i\phi}\cos \theta}&
\dfrac{e^{2i\phi} - e^{i\omega_{x+1}}\sin \theta e^{-i\omega_{x}}\sin \theta}{-e^{i\phi}\cos \theta}
\end{bmatrix} \\
&=
\begin{bmatrix}
e^{-i\phi} \cos \theta &
e^{i (\omega_{x+1} - \phi)} \sin \theta \\
e^{-i (\omega_{x} + \phi)}\sin \theta &
-\dfrac{e^{i\phi} - e^{i (\omega_{x+1} - \omega_{x} - \phi)}\sin^{2} \theta}{\cos \theta}
\end{bmatrix} \\
&=
\begin{bmatrix}
e^{-i\phi}\cos \theta & e^{i\alpha_{x+1}}\sin \theta \\
e^{-i\alpha_{x+1}}\sin \theta & -e^{i\phi}\cos \theta 
\end{bmatrix}.
\end{align*}
These transfer matrices $D_{x}^{+}$ and $D_{x}^{-}$ are unitary matrices. 
By Eq.(3.1) in Lemma 3.1, if transfer matrices are unitary matrices, 
the norm of $\Psi(x)$ is independent of position $x$.
Therefore, the conclusion $\Gamma (\Psi) \in \mathcal{M}_{unif}$
can be derived from Lemma 3.1. 
\qed \\

Ohno \cite{Ohno2016, Ohno2017} investigated unitary equivalent classes of one-dimensional QWs.
Using the method in \cite{Ohno2017}, we can also prove Theorem 3.2. However, 
compared with his method, our method is easier to show our result.

\section{Application to cycles}
In this section, we consider two-state QWs on a cycle $C_{2N}$ with $2N$ vertices for $N \in \mathbb{N}$. 
Here a cycle $C_{m} = (V, E)$ with $m \in \mathbb{N}$ is defined by the set of vertices, 
\begin{align*}
V = \{x \in \mathbb{Z} / m\mathbb{Z} \}, 
\end{align*}
and the set of edges, 
\begin{align*}
E = \{ (x, x + 1), (x + 1, x) : x \in V \}.
\end{align*}
We define QWs on a cycle $C_{m}$ whose coin matrix at position $x$ is given by $U_{x}$.
As in the case of $\mathbb{Z}$, the time evolution of the walk is determined by
\begin{align*}
\Psi_{n+1}(x) \equiv P_{x+1} \Psi_{n}(x+1) + Q_{x-1} \Psi_{n}(x-1) \hspace{2.em} (x \in \mathbb{Z} / m \mathbb{Z}) .
\end{align*}
The time evolution operator in the cycle case is given by $U_{c}^{(s)}$ like $U^{(s)}$ in the case of $\mathbb{Z}$.
Then we see that $\Psi_{n+1} = U_{c}^{(s)} \Psi_{n}$ $(n \geq 0)$, where 
\begin{align*}
\Psi_{n}
= {}^{T}[\hspace{.3em}...\hspace{.4em}, \Psi^{L}_{n}(-1),  \Psi^{R}_{n}(-1), \Psi^{L}_{n}(0),  \Psi^{R}_{n}(0),
 \Psi^{L}_{n}(+1),  \Psi^{R}_{n}(+1), \hspace{.4em}...\hspace{.3em}]. 
\end{align*}
We should remark that $\prod_{x=1}^{m} D_{x}^{+} = I_{2}$ folds for our cycle $C_{m}$ case, 
if we can replace $U^{(s)}$ with $U^{(s)}_{c}$ in Theorem 3.2, where $I_{2}$ is the $2\times2$ identity matrix. 
Using this remark, we can get the following result.
\begin{pro}
We consider the QWs on a cycle $C_{2N}$ with $N \in \mathbb{N}$ whose coin matrix at position $x$ is given by
\begin{align*}
U_{x} = 
\begin{bmatrix}
\cos \theta & e^{i\omega_{x}}\sin \theta \\
e^{-i\omega_{x}}\sin \theta & -\cos \theta \\ 
\end{bmatrix}
\hspace{2.em}(\omega_{x} \in [0, 2\pi), \theta \in (0, 2\pi) ),
\end{align*}
where $\omega_{x}-\omega_{x-1} =  \frac{2\pi}{N} \hspace{.3em}(\hspace{-.6em} \mod\hspace{.2em}2\pi)$
 for all $x \in \mathbb{Z}$ and $\theta \neq \frac{\pi}{2}$, $\theta \neq \frac{3\pi}{2}$. 
This model has a uniform stationary measure. 
\end{pro}
{\bf Proof.} In this case, we obtain
\begin{align*}
D_{x}^{+} =\begin{bmatrix}
e^{i \frac{\pi}{N}}\cos \theta & e^{i\alpha_{x}}\sin \theta \\
e^{-i\alpha_{x}}\sin \theta & -e^{-i \frac{\pi}{N}}\cos \theta 
\end{bmatrix},
\end{align*}
where $\alpha_{x} = \frac{\pi}{N} + \omega_{x-1} = \omega_{x}-\frac{\pi}{N}
 \hspace{.2em}(\hspace{-.6em} \mod\hspace{.2em}2\pi)$.
By definition of $\alpha_{x}$, we get
\begin{align}
\alpha_{x+1} - \alpha_{x} = 
\left(\frac{\pi}{N} + \omega_{x}\right) - \left(\frac{\pi}{N} + \omega_{x-1}\right) = 
\frac{2\pi}{N} \hspace{2.em} (\hspace{-.9em} \mod \hspace{.2em}2\pi)
\end{align}
and
\begin{align}
\alpha_{x+1} - \frac{\pi}{N} = \alpha_{x} + \frac{\pi}{N}
\hspace{2.em} (\hspace{-.9em} \mod \hspace{.2em}2\pi).
\end{align}
Thus we see
\begin{align*}
D_{x+1}^{+} D_{x}^{+} = 
\begin{bmatrix}
e^{i \frac{\pi}{N}}\cos \theta & e^{i\alpha_{x+1}}\sin \theta \\
e^{-i\alpha_{x+1}}\sin \theta & -e^{-i \frac{\pi}{N}}\cos \theta
\end{bmatrix}
\begin{bmatrix}
e^{i \frac{\pi}{N}}\cos \theta & e^{i\alpha_{x}}\sin \theta \\
e^{-i\alpha_{x}}\sin \theta & -e^{-i \frac{\pi}{N}}\cos \theta
\end{bmatrix}
=
\begin{bmatrix}
e^{i  \frac{2\pi}{N}} & 0 \\
0 & e^{-i  \frac{2\pi}{N}}  \\ 
\end{bmatrix}.
\end{align*}
The second equality derived from the Eqs.(4.2) and (4.3). 
Therefore, we have
\begin{align*}
 \prod_{x=1}^{2N} D_{x}^{+} 
= 
\begin{bmatrix}
e^{i  \frac{2\pi}{N}} & 0 \\
0 & e^{-i  \frac{2\pi}{N}}  \\ 
\end{bmatrix} ^{N}
= I_{2}.
\end{align*}
Thus, the desired conclusion can be obtained.
\qed

\section{Periodicity of coins of QW on the line}
In this section, we consider the periodicity of the sequence of coins $\{U_{x}\}_{x \in \mathbb{Z}}$ which 
determines QW. Put $N \in \mathbb{N}$. If $U_{x + N} = U_{x}$ for $x \in \mathbb{Z}$, we say that 
$\{U_{x}\}_{x \in \mathbb{Z}}$ has $N$ period. If not, $\{U_{x}\}_{x \in \mathbb{Z}}$ has no period.

By Theorem 3.2, we get the following result.
\begin{cor}
We consider QWs on the line with $\{U_{x}\}_{x \in \mathbb{Z}}$ in which for each $x \in \mathbb{Z}$, 
satisfying $U_{x} \in \mathcal{C}_{\phi}$ with $\phi \in [0, 2\pi)$, i.e.,  
\begin{align*}
U_{x} = 
\begin{bmatrix}
\cos \theta & e^{i\omega_{x}}\sin \theta \\
e^{-i\omega_{x}}\sin \theta & -\cos \theta \\ 
\end{bmatrix}
\hspace{2.em}(\omega_{x} \in [0, 2\pi), \theta \in (0, 2\pi)),
\end{align*}
where $\omega_{x}-\omega_{x-1} =  \frac{2\pi}{N} \hspace{.3em}(\hspace{-.6em} \mod\hspace{.2em}2\pi)$
 for all $x \in \mathbb{Z}$.
In this setting, we have \\
Case 1. If we choose $\phi = \frac{1}{N} \times \pi$ with $N \in \mathbb{N}$, then $\{U_{x}\} _{x \in \mathbb{Z}}$ has $N$ period. \\
Case 2. If we choose $\phi = a\pi$ where a is an irrational number, then $\{U_{x}\} _{x \in \mathbb{Z}}$ has no period. 
\end{cor}
We give an example corresponding to case 1 in Corollary 5.1. In this case, we put $\phi = \frac{1}{3}\times\pi$ (i.e., $N = 3$) and $\omega_{0} = 0$. 
Then $\{U_{x}\}_{x \in \mathbb{Z}}$ can be written as 
\begin{align*}
U_{x} = 
\left\{
\begin{array}{l}
U_{0} \hspace{.6cm} (x = 0 \hspace{.2cm} \mbox{mod}\hspace{.2cm} 3), \\
U_{1} \hspace{.6cm} (x = 1 \hspace{.2cm} \mbox{mod}\hspace{.2cm} 3), \\
U_{2} \hspace{.6cm} (x = 2 \hspace{.2cm} \mbox{mod}\hspace{.2cm} 3), 
\end{array}
\right.
\end{align*}
where
\begin{align*}
U_{0}= 
\begin{bmatrix}
\cos\theta & \sin \theta  \\
\sin \theta & -\cos \theta  \\ 
\end{bmatrix}, \hspace{1.em}
U_{1} = 
\begin{bmatrix}
\cos\theta & e^{\frac{2}{3}\pi i}\sin \theta  \\
e^{-\frac{2}{3}\pi i}\sin \theta & -\cos \theta  \\ 
\end{bmatrix}, \hspace{1.em}
U_{2} = 
\begin{bmatrix}
\cos\theta & e^{\frac{4}{3}\pi i}\sin \theta  \\
e^{-\frac{4}{3}\pi i}\sin \theta & -\cos \theta  \\ 
\end{bmatrix}.
\end{align*}

\section{RW case}
This section is devoted to stationary measures of RWs on $\mathbb{Z}$. 
For more details on RWs, see \cite{Schinazi1999}.
The RW is determined by a sequence of the classical counterpart of coins 
$\{p_{x}\}_{x \in \mathbb{Z}}$ with $p_{x} \in [0,1]$.
That is to say, the time evolution of the RW is defined by 
\begin{align}
\mu_{n}(x) = p_{x+1}\mu_{n-1}(x+1) + q_{x-1}\mu_{n-1}(x-1),
\end{align}
where $\mu_{n}(x)$ is the measure for the RW at time $n$ and position $x$, 
and $q_{x} = 1 - p_{x}$.
Here we introduce the transition matrix $P^{(s)}$ of the RW as follows
\begin{align*}
P^{(s)}
= 
\left[ 
\begin{array}{ccccccc}
\ddots & \vdots & \vdots & \vdots & \vdots & \vdots & \vdots \\
\cdots & q_{-2} & 0 & p_{0} & 0 & \cdots & \cdots \\
\cdots & 0 & q_{-1}& 0 & p_{1} & 0 & \cdots \\
\cdots & \cdots & 0 & q_{0} & 0 & p_{2} & \cdots \\
\cdots & \vdots & \vdots & \vdots & \vdots & \vdots & \ddots 
\end{array} 
\right].
\end{align*}

Let $\mathcal{M}(P^{(s)})$ be the set of all measures of the RW.
Furthermore, $\mathcal{M}_{s}(P^{(s)})$ denotes the set of all stationary measures of 
the RW, i.e., 
\begin{align*}
\mathcal{M}_{s}(P^{(s)}) = \Big\{\mu \in \mathcal{M}_{s}(P^{(s)}) : P^{(s)} \mu = \mu \Big\}
\end{align*}
In this setting, we obtain the following result.
\begin{pro}
Assume that $\mathcal{M}_{unif} \subset \mathcal{M}_{s}(P^{(s)})$.
Then, $\{p_{x}\}_{x \in \mathbb{Z}}$ has 1 period (i.e., $p_{x}$ is a constant) or 2 period.
\end{pro}
{\bf Proof.} 
The assumption implies that there exists a uniform stationary measure $\mu_{n}(x) = c \hspace{.2 em}( > 0)$
for any $n = 0, 1, 2, ...$ and $x \in \mathbb{Z}$. Thus, we see that Eq.(6.4) becomes 
\begin{align*}
c = p_{x+1} c + (1 - p_{x-1})c,
\end{align*}
since $q_{x-1} = 1-p_{x-1}$. Therefore we have 
\begin{align*}
p_{x-1} = p_{x+1} \hspace{2. em} (x \in \mathbb{Z})
\end{align*}
and the proof is complete.
\qed \\

Table 1 summarizes our results on Corollary 5.1 (QW case) and Proposition 6.1 (RW case) and clarifies the difference between them, where ``$\circ$'' means there exists a sequence of coins with $n$ period for $n = 0, 1, 2, \ldots, \infty$ and ``$\times$'' means ``otherwise". Here ``$\infty$ period" is equivalent to ``no period".

\begin{table}[htb]
\centering
\caption{}
\scalebox{1.3}[1.5]{
\begin{tabular}{| c | c | c | c | c | c | c |} \hline
The number of the period & 1 & 2 & 3 & 4 & $\cdots$ & $\infty$ \\ \hline
RW & $\circ$ & $\circ$ & $\times$ & $\times$ & $\cdots$ & $\times$ \\ \hline
QW & $\circ$ & $\circ$ & $\circ$ & $\circ$ & $\cdots$ & $\circ$ \\ \hline
\end{tabular}
}
\end{table}

\section{Summary}
In this paper, we studied stationary measures of discrete-time two-state QWs. 
We presented models having the uniform stationary measure on $\mathbb{Z}$ and cycles. 
Moreover, 
considering the sequence of coins, 
we found out an interesting difference between QWs and RWs. 
As a future work, it would be fascinating to extend our result and observation to QWs and RWs on general graphs.

\par
\
\par\noindent
{\bf Acknowledgments.} 
We are grateful to Hiromichi Ohno for his helpful comments.
Y. I. was supported by the Grant-in-Aid for Young Scientists (B) of Japan Society for the Promotion of Science (Grant No. 16K17652).

\begin{small}
\bibliographystyle{plain}

\end{small}

\end{document}